\documentclass[conference]{IEEEtran}
\IEEEoverridecommandlockouts
\usepackage{cite}
\usepackage{amsmath,amssymb,amsfonts}
\usepackage{algorithmic}
\usepackage{graphicx}
\usepackage{textcomp}
\usepackage{hyperref}
\usepackage{xcolor}
\usepackage{tabularx}
\usepackage{multirow}
\usepackage{subcaption}

\def\BibTeX{{\rm B\kern-.05em{\sc i\kern-.025em b}\kern-.08em
    T\kern-.1667em\lower.7ex\hbox{E}\kern-.125emX}}
\begin{document}

\title{A Visual-Analytical Approach for Automatic Detection of Cyclonic Events in Satellite Observations\\

\thanks{This work is in collaboration with ISRO-SAC approved by ISRO-IISc STC and ARTPARK IISc.}
}

\author{\IEEEauthorblockN{Akash Agrawal}
\IEEEauthorblockA{
\emph{Indian Institute of Science} \\
Bangalore, India \\
akashagrawa1@iisc.ac.in}
\and
\IEEEauthorblockN{Mayesh Mohapatra }
\IEEEauthorblockA{
\emph{Indian Institute of Science} \\
Bangalore, India \\
mayeshm@iisc.ac.in}
\and
\IEEEauthorblockN{Abhinav Raja}
\IEEEauthorblockA{
\emph{BITS Pilani} \\
Goa, India \\
f20191041@goa.bits-pilani.ac.in}
\and
\IEEEauthorblockN{Paritosh Tiwari}
\IEEEauthorblockA{
\emph{Indian Institute of Science} \\
Bangalore, India \\
paritosht@iisc.ac.in}
\and
\hspace{10pt}
\IEEEauthorblockN{Vishwajeet Pattanaik}
\hspace{10pt}
\IEEEauthorblockA{
\hspace{10pt}
\emph{Indian Institute of Science} \\
\hspace{10pt}
Bangalore, India \\
\hspace{10pt}
vishwajeetp@iisc.ac.in}
\and
 \hspace{10pt}
 \IEEEauthorblockN{Neeru Jaiswal}
 \hspace{10pt}
 \IEEEauthorblockA{
  \hspace{10pt}
 \emph{
 SAC, ISRO} \\
  \hspace{10pt}
 Ahmedabad, India \\
  \hspace{10pt}
 neeru@sac.isro.gov.in}
 \and
 \hspace{20pt}
 \IEEEauthorblockN{Arpit Agarwal}
 \hspace{20pt}
 \IEEEauthorblockA{
 \hspace{20pt}
 \emph{
 SAC, ISRO} \\
 \hspace{20pt}
 Ahmedabad, India \\
 \hspace{20pt}
 arpitagarwal@sac.isro.gov.in}
 \and
 \hspace{20pt}
 \IEEEauthorblockN{Punit Rathore}
 \hspace{20pt}
\IEEEauthorblockA{
\hspace{20pt}
\emph{Indian Institute of Science} \\
\hspace{20pt}
Bangalore, India \\
\hspace{20pt}
prathore@iisc.ac.in}
}
\maketitle
\begin{abstract}
Estimating the location and intensity of tropical cyclones holds crucial significance for predicting catastrophic weather events. In this study, we approach this task as a detection and regression challenge, specifically over the North Indian Ocean (NIO) region where best tracks location and wind speed information serve as the labels. The current process for cyclone detection and intensity estimation involves physics-based simulation studies which are time-consuming, only using image features will automate the process for significantly faster and more accurate predictions. While conventional methods typically necessitate substantial prior knowledge for training, we are exploring alternative approaches to enhance efficiency.   
\end{abstract}


\section{Introduction}
Cyclones are giant, spiraling storms that can form over oceans. They typically form over warm ocean waters near the equator. India is exposed to nearly 10 percent of the world’s tropical cyclones. This research aims to explore automatic event detection in remote sensing data, specifically tropical cyclones, using a supervised learning approach and generating a highly interpretable visual representation of said event for monitoring and analysis. The current work is to develop modules for both detection and intensity estimation, which can be integrated in future to deliver an end to end pipeline for development of a visual analytic tool for cyclone detection. We hope to enable rapid progress in remote sensing applications and facilitate data-driven approaches to solve global-scale challenges such as disaster prevention, and specifically help meet ISRO's requirement of faster inferencing as opposed to the current NWP models for automatic cyclone detection.

\subsection{MOTIVATION}
Currently, multiple models exist in the domain of weather forecasting, but most of these models range from physics-based models, statistics-based models, and Numerical Weather Prediction (NWP) models to differential equation-based models. As we increase the complexity of the tasks, these models take up a lot of time and are not efficient. Additionally, these models rely on surface-level data such as wind speed, surface pressure, and wind direction as well along with satellite imagery, this surface-level data is not so easily obtained. Currently, NWP models are being used at ISRO for weather prediction, which requires extensive resources and computation time. 

\subsection{OUR CONTRIBUTION}
This research aims to focus specifically on cyclone detection, intensity estimation and related aspects using only image input and data-driven approaches and will lead to faster inference time and automate the process as opposed to current NWP models being utilized at SAC.
In context to algorithm development, a novel two stage detection and intensity estimation module is proposed. In the first level detection we try to localise the cyclone over an entire image as captured by INSAT3D over the NIO (North Indian Ocean). For the intensity estimation task, we propose a CNN-LSTM network, which works on the cyclone centred images, utilising a ResNet-18 backbone, by which we are able to capture both temporal and spatial characteristics.

\section{RELATED WORK}

\subsection{Tropical Cyclone Geolocation Detection}
Cyclone tracking and forecasting have traditionally relied on a combination of satellite imagery, atmospheric data, and historical weather patterns. Objective Detection of Center of Tropical Cyclone in Remotely Sensed Infrared Images~\cite{a1} uses the fluxes of the gradient vectors of brightness temperature (BT) to determine the point around which they are converging. 
Image processing-based approaches to localize the center of tropical cyclones (TCs) have also been developed such as in~\cite{a2}, where the authors propose an algorithm by extracting the spiral features within TCs in infrared imagery taken via a geo-stationary satellite. A novel algorithm for detecting center of tropical cyclone in satellite infrared images~\cite{a3} utilises edge detection and density matrix to determine the center of TC. These techniques work on already centered images which are initialised by a guess center, and applicability to multiple cyclone scenarios is not discussed. 
Identification of tropical cyclone center based on satellite images via deep learning techniques~\cite{a4} uses YOLOv4 over non cyclone centered images spanning a wider area, making the detection challenging but practical. The work also demonstrates there is higher positional error in detection for lower intensity classes.
A center location algorithm for tropical cyclone in satellite infrared images~\cite{a5} considers  non-centred images and utilises OS(One shot detectors like YOLOv3) for initial detection and subsequent image processing (IP)  over cyclone centered images to achieve significant improvement in results, although not considering cyclones at lower intensity (Tropical depression), which are often difficult to detect.
Determination of low-intensity tropical cyclone centers in geostationary satellite images using a physics-enhanced deep-learning model~\cite{a6} specifically tries to address the problem of large errors in lower intensity class detection using both satellite image sequences and historical TC CLIPER information as inputs.
However, most of the above do not discuss results on explicit Non-Cyclonic data, where precision is also crucial during actual deployment.
\vspace{-2pt}

\subsection{Tropical Cyclone Intensity Estimation}
The Dvorak Technique~\cite{b1} has been a standard to measure cyclone intensities since the 1970s and has been used as a reliable source for predicting intensities based on the shape of the cyclone. 

These methods do not leverage the advantages of using ML models for predicting cyclone intensities. Convolutional neural networks (CNNs) have recently been used for cyclone tracking and forecasting~\cite{b3},\cite{b4},\cite{b10},\cite{b5}. In~\cite{b3}, the authors use interpolation and data-augmentation techniques to enhance the temporal resolution for cyclone images and then use these augmented images on a lightweight model to classify cyclones. In~\cite{b4}, the authors propose a two-stage process for intensity categorization into 3-bins of cyclone classes and then use class-wise regressors on each of these bins to assign the intensity to these cyclones. This would lead to a resource-heavy algorithm for a multi-class classification problem. For intensity estimation, ~\cite{b10} uses a CNN model and a regression based approach. In~\cite{b5}, the proposed model makes use of a cyclone classification network fused with Harlick features, MB-LBP, and geometric features. In~\cite{b6}, an SVM based classifier, that uses GLCM Harlick Features is proposed. 
In ~\cite{b14}, the authors integrate Distance-Consistency (DC) and Rotation-invariance (RI) features, this fusion significantly improves the accuracy of cyclone intensity prediction. Additionally, the authors employ a transformer model to effectively capture and aggregate temporal correlations present in sequential cyclone images, further refining the estimation process. This makes training a huge task.

\section{CONTRIBUTION AND RESULTS}

\subsection{DATASET}
The data was downloaded through MOSDAC ((Meteorological \& Oceanographic Satellite Data Archival Centre) portal for the archival of satellite data. The dataset used contained images taken by ISRO’s INSAT-3D IR Imager Satellite~\cite{b11}.The ground resolution at the sub-satellite point is nominally 4km x 4km for TIR bands.
We obtain a total 18330 h5 files based on the event information shared in IMD Best Tracks Data. Each image is a TIR-1(10.2 - 11.2 µm) IR Image of size 3207x3062, taken by INSAT-3D satellite at 30 minutes intervals. During the transformation process, the original brightness temperature is normalized to the grayscale image range of 0 to 255 by considering brightness temperature range from 180K to 310K.
In addition to the image dataset, Best-Track data for Indian TCs was retrieved from the India Meteorological Department (IMD). Together the image and best-track data, covers all TC events from 2013 to 2023. 
 \begin{figure}[!ht]
    \centering
    \includegraphics[width=1\linewidth]{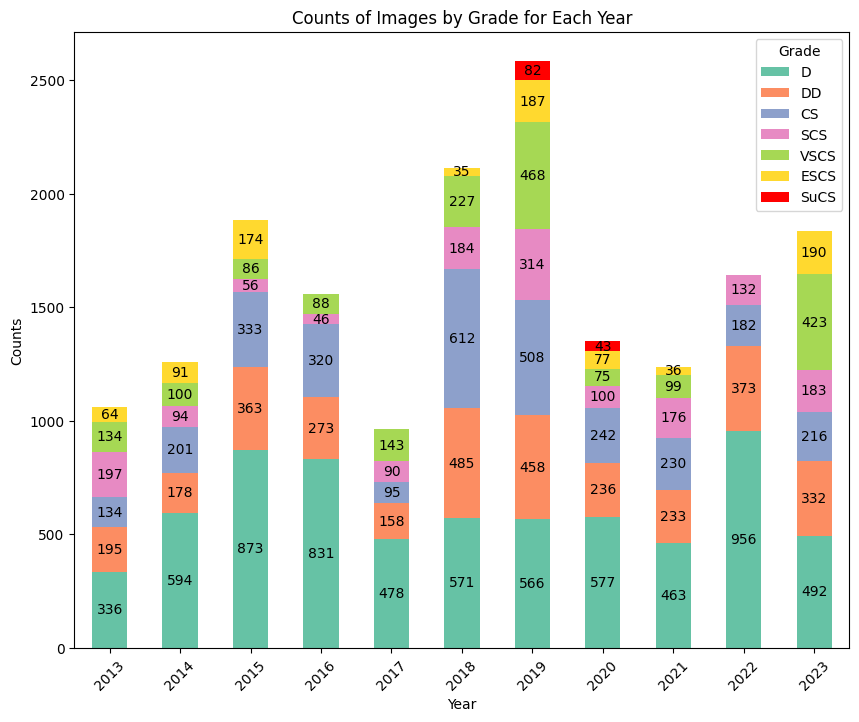}
    \caption{Year-wise Class Distribution of Cyclones (2013-2023).}
    \label{fig:classes_plot}
\end{figure}

 The classification adheres to the IMD's wind-speed-based taxonomy, namely, Depression (D; 17-27 knots), Deep Depression (DD; 28-33 knots), Cyclonic Storm (CS; 34-47 knots), Severe Cyclonic Storm (SCS; 48-63 knots), Very Severe Cyclonic Storm (VSCS; 64-89 knots), Extremely Severe Cyclonic Storm (ESCS; 90-119 knots), and Super Cyclonic Storm (SuCS; wind speeds above 120 knots). 
We have used linear interpolation to generate labels for half-hourly images, as best track data was available at 3hr intervals.

\subsubsection{\textbf{Cyclone Geolocation Dataset Creation}}
A special scheme was designed to extract relevant regions from (North Indian Ocean) NIO. From each h5 file we have extracted two relevant regions, corresponding to Arabian Sea (AS) and Bay of Bengal (BOB). This helps us to capture region specific features, for more accurate detections. 
Data within Latitude Ranges from: 0°S-35°N, Longitude Ranges from: 45°E-82.2°E for AS and 75-112.2°E for BOB was extracted from the satellite images, to avoid the models learning features that are not relevant to the North Indian Ocean TC events. The ranges were selected to generate uniform square sized shape images of pixel dimensions 1035x1035 to avoid any spatial deformation before resizing. Figure~\ref{fig:dataset_extraction} illustrates the dataset preprocessing for the Cyclone Detection model. 

\subsubsection{\textbf{Cyclone Intensity Estimation Dataset Creation}}
\begin{figure}[!ht]
    \centering
    \includegraphics[width=1\linewidth]{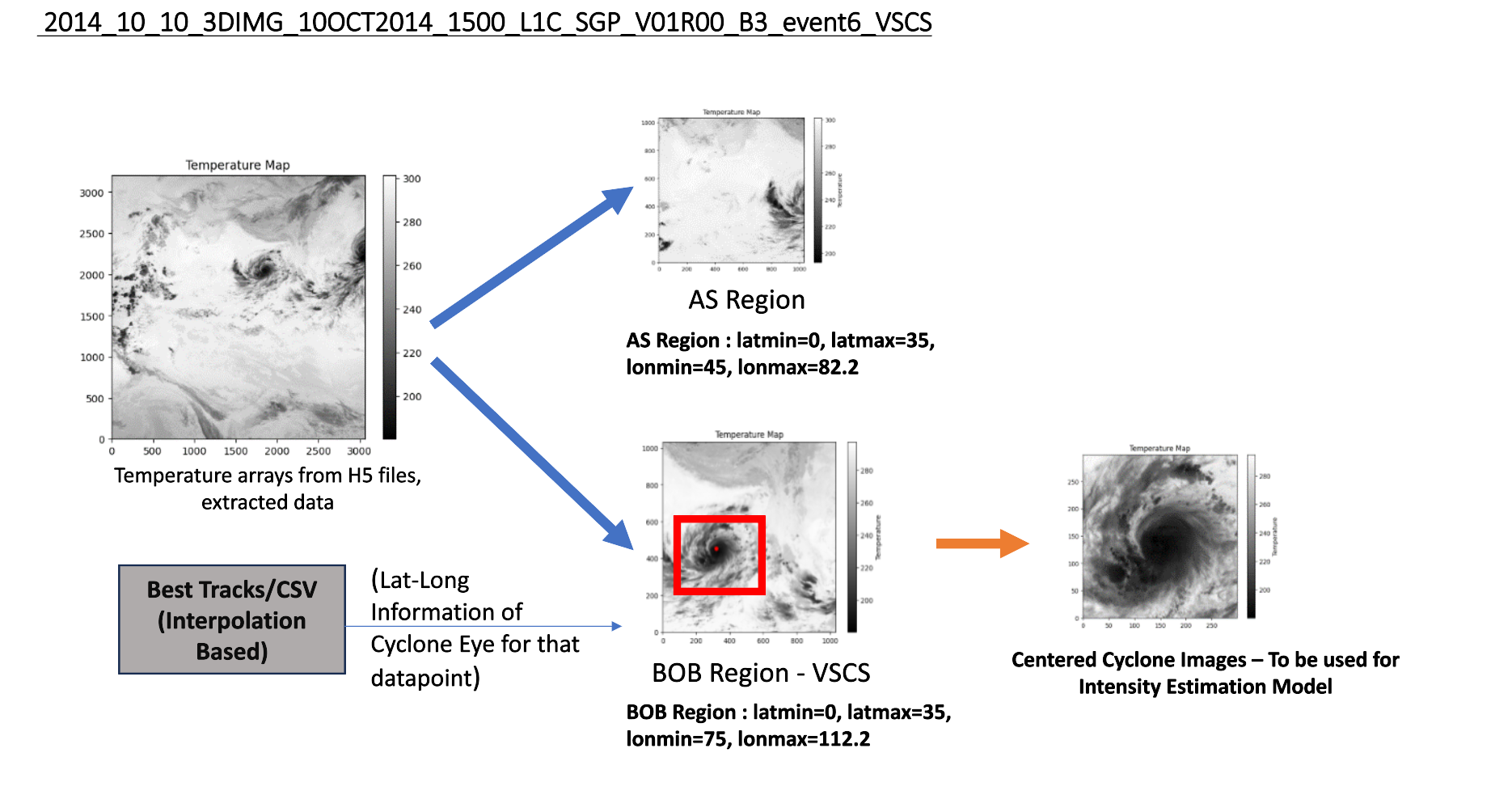}
    \caption{TC Dataset Extraction Scheme}
    \label{fig:dataset_extraction}
\end{figure}

Using INSAT 3D satellite specific projection information and best track latitude longitude information, we extracted the cyclone centric images to be used for developing the intensity estimation module. We have carefully avoided the use of data in which cyclone landfall has occurred in testing, as the cloud structure is different in such scenarios and will not be accurate representation of TCs over sea. Figure~\ref{fig:dataset_extraction} illustrates the extraction process which used for creating the dataset used as input for the Cyclone Intensity model. 

\subsubsection{\textbf{No Cyclone (NC) Data for Testing Purpose}}
For generating NC data, we followed 2 different schemes. The extraction process was similar for cyclonic data as described above and images of size 1035 x 1035 were used.\\
a) \textbf{Type I NC Data}: Images from pre-48hr from cyclone start and upto post 48hr after cyclone ends, 1499 such images were randomly sampled.\\
b) \textbf{Type II NC Data}: Images between pre 162hr-48hr from cyclone start and between post 48hr-162hr after cyclone ends, 1500 such images were randomly sampled.\\

\subsection{MODEL DEVELOPMENT}
\subsubsection{\textbf{TC Geolocation Detection}}
We model the cyclone geolocation estimation as an object detection problem. Currently there are single shot detectors like the YOLO series of models ~\cite{b8} and two stage detectors like Fast RCNN~\cite{b9}) as choices for object detection modelling. Identification of tropical cyclone center based on satellite images via deep learning techniques~\cite{a4} draws a comparision between 6 different object detection models and concludes Yolov4 performs the best in terms of all performance metrics including mean lat-lon error, with RFB (Receptive Field Block) coming second. The single shot methods predict the bounding boxes and have faster inference capabilities, directly returning the object locations and confidence estimates. Drawing motivation from above, we have preferred to use the latest state of the art SSD in the YOLO series, YoloV7~\cite{b16}) and YoloV8~\cite{b17}) in this study.

\begin{figure}[!ht]
    \centering
    \includegraphics[width=1\linewidth]{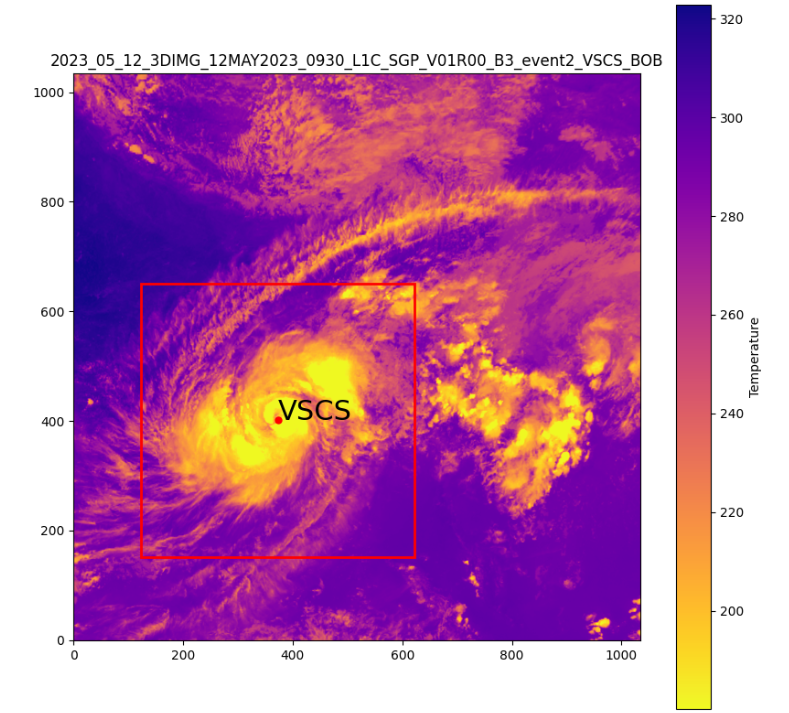}
    \caption{TC MOCHA - Bounding Box Plot}
    \label{fig:YOLO_Label}
\end{figure}

\begin{figure}[!ht]
    \centering
    \includegraphics[width=1\linewidth]{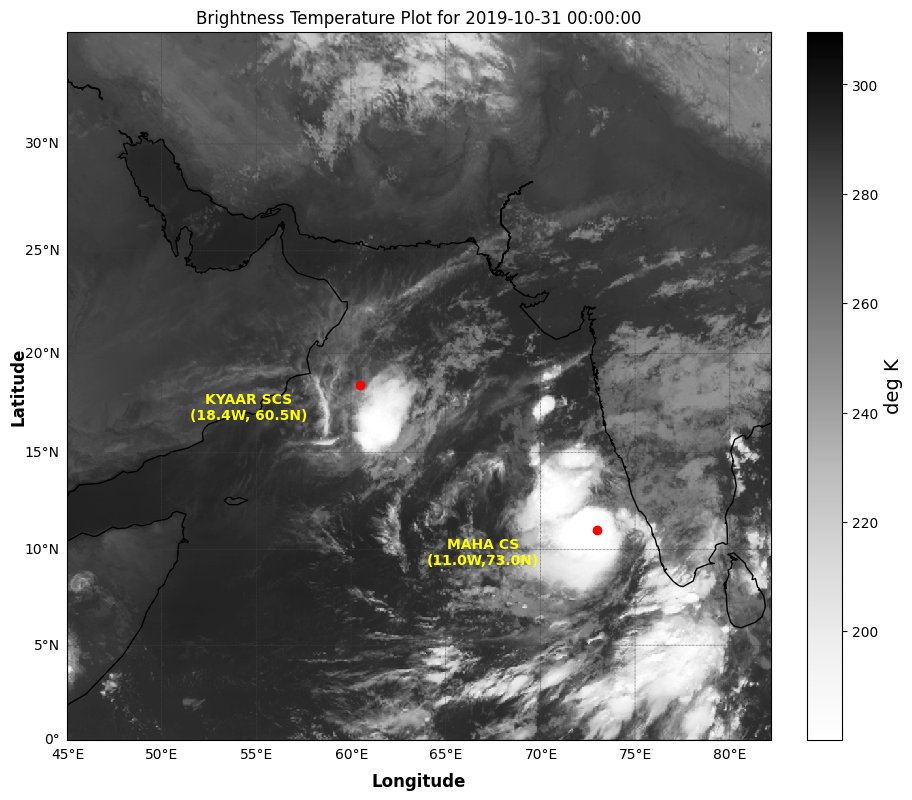}
    \caption{TC MAHA and KYAAR in AS, red dot denotes
    best tracks location of cyclones}
    \label{fig:Multicyclone}
\end{figure}

Using the Lat-Lon information of cyclone center from best tracks, we created the labels in YOLO format corresponding to the AS and BOB images. An example bounding box plot over a temperature map for MOCHA cyclone(2023) is shown in Figure~\ref{fig:YOLO_Label}. There were also instances when two cyclones occur in the same basin, as shown in Figure~\ref{fig:Multicyclone}. This multi-cyclonic event was part of the test dataset, to check the efficacy of model in detecting multiple cyclones. We experimented with different box sizes based on cyclone intensity starting from 500 x 500 to 200 x 200, with the idea to capture all important features for all classes of cyclones. The final configuration selected for classes D,DD,CS,SCS,VSCS was 400 x 400 and classes ESCS,SuCS was 300 x 300, optimised for precision and recall, discussed in the results section. The intensity class information was also encoded in the label files.
For training and validation, we have considered data from years 2013-2022, and sampled it in a stratified way, to maintain the class distribution in train and validation sets. 

\subsubsection{\textbf{TC Intensity Estimation}}
\begin{figure}[!ht]
    \centering
    \includegraphics[width=1\linewidth]{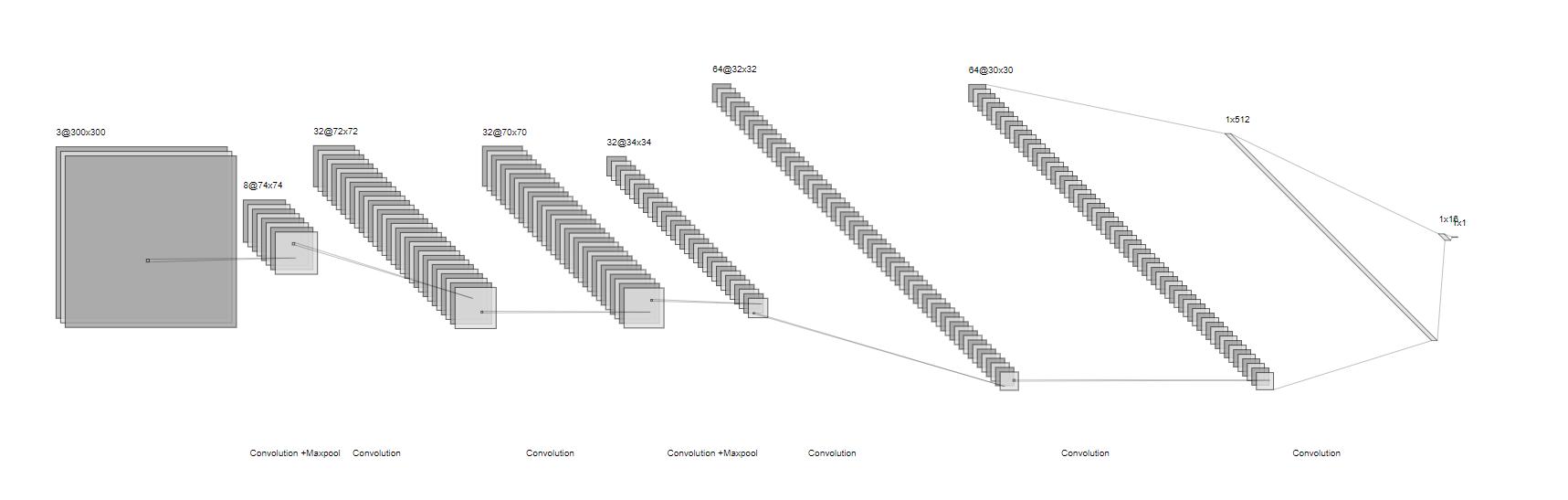}
    \caption{The CNN architecture used for TC Intensity Estimation using single image (M1).}
    \label{fig:SIR_architecture}
\end{figure}

For TC intensity estimation, we are using cyclone centered images. Intensity of a cyclone can be calculated as the near-surface maximum wind speed around the circulation centre. The wind speed values and intensity classes were taken from IMD best tracks. There are several modelling approaches, one involves using single IR image to determine intensity as in CNN regression based architecture, Maskey, M. et al (2020) ~\cite{b10} or a combination of classifier and regression models, Zhang,CJ. et al. (2021)  ~\cite{b4} , and other involves using sequences of past images with the current image. We have explored the below approaches. 
\paragraph{\textbf{Single image regression model (M1)}}\label{AA}
The single image regression model we used is based on a custom CNN architecture.  We have used input dimensions of 300 x 300 image size, and two max pooling layers and six convolutional layers. The architecture is depicted in the Figure ~\ref{fig:SIR_architecture}.

\begin{figure}[!ht]
    \centering
    \includegraphics[width=1\linewidth]{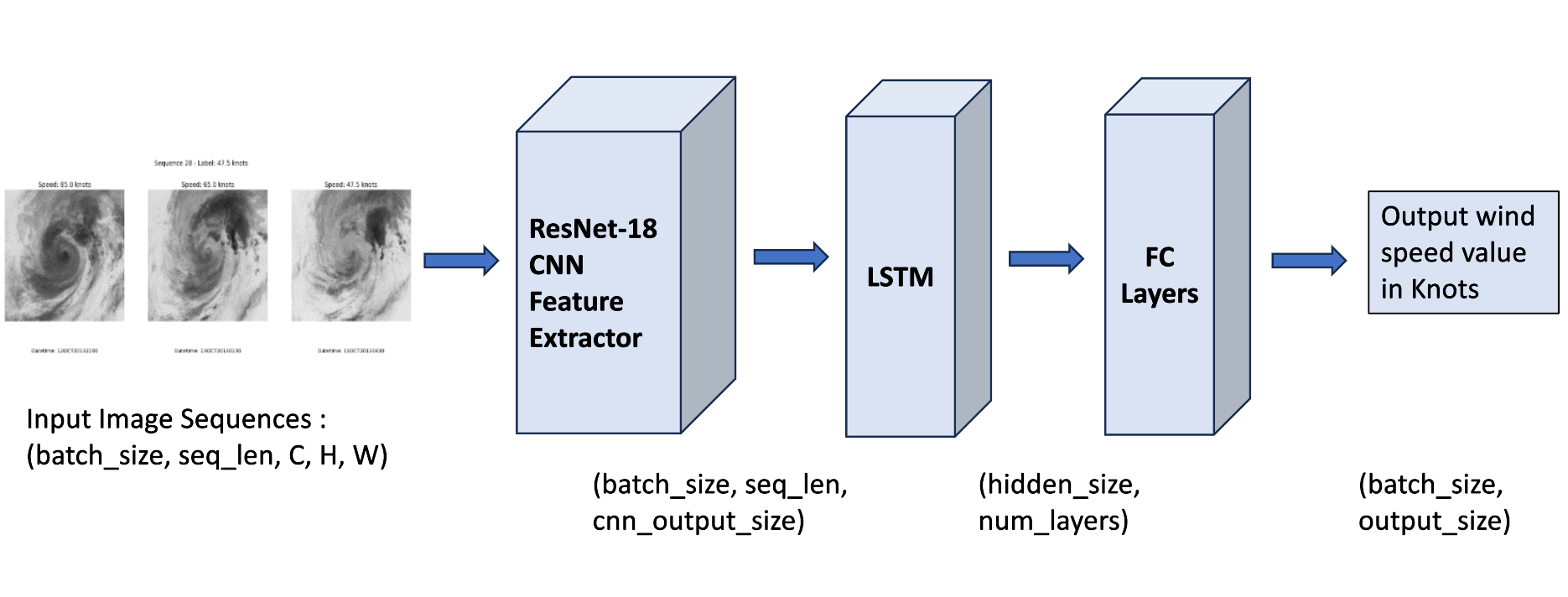}
    \caption{Framework of CNN-LSTM Model (M2)}
    \label{fig:CNN_LSTM}
\end{figure}

\begin{figure}[!ht]
    \centering
    \includegraphics[width=1\linewidth]{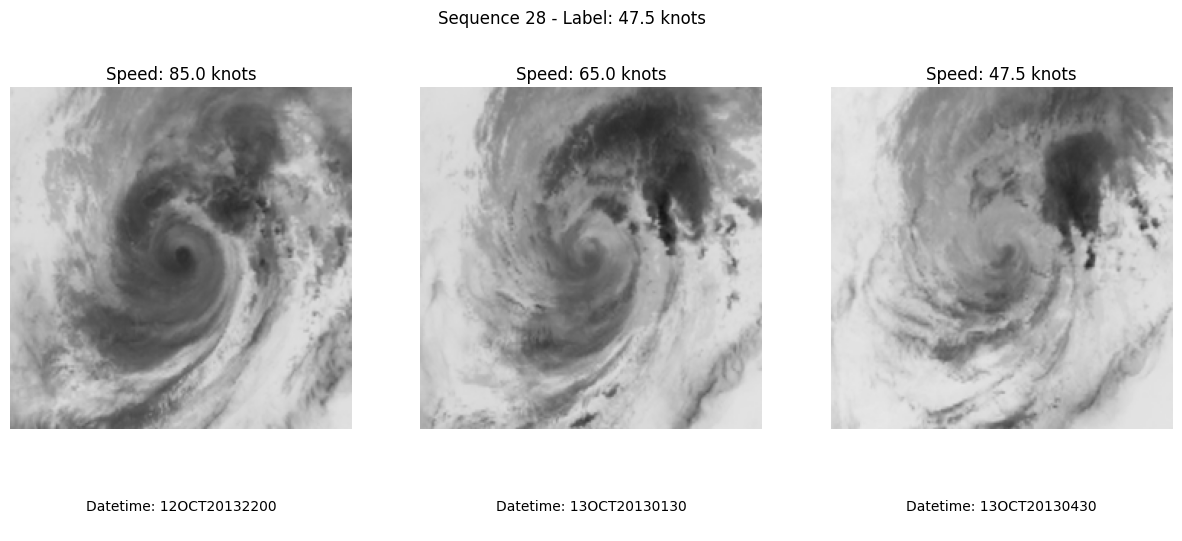}
    \caption{A training sequence for CNN- LSTM model : Cyclone PHAILIN in its weakening phase, 13OCT2013, 0430. }
    \label{fig:training_sequence}
\end{figure}

\paragraph{\textbf{Sequential CNN-LSTM Model (M2)}}\label{AA}
The Custom CNN-LSTM network that uses historic images along with the current image, uses ResNet-18 as a backbone. The block diagram for the architecture is depicted in Figure ~\ref{fig:CNN_LSTM}. We have experimented with different sequence lengths, intervals (between two consecutive images) and duration (the time up-till which the historic images are considered). One such sequence is depicted in Figure~\ref{fig:training_sequence}, for duration 6 hr and interval 3hr. To handle missing images (as all half-hourly timestamps are not available on MOSDAC consistently), we have also included a buffer of time interval/6, to construct sequences.

\subsubsection{\textbf{Model Explainability}}
Eigen Class Activation Mapping (Eigen-CAM) Muhammad et. al(2020)~\cite{b15}, uses the principle components of learned representations from the convolutional layers to produce a coarse localisation map, not relying on the backpropogation of gradients of any target concept flowing into the final convolutional layer, as in GradCAM. They also demonstrate that empirically there was a 12\% improvement over the other state of the art methods such as GradCAM and GradCAM++.

\begin{figure}
	\centering
	\begin{subfigure}{0.45\linewidth}
		\includegraphics[width=\linewidth]{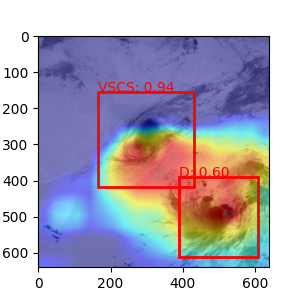}
		\caption{}
		\label{fig:subfigMC}
	\end{subfigure} \hspace{0.5mm}
	\begin{subfigure}{0.45\linewidth}
		\includegraphics[width=\linewidth]{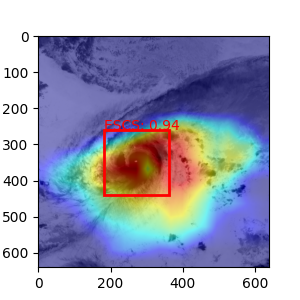}
		\caption{}
		\label{fig:subfigSC}
	\end{subfigure}

	\caption{Eigen-CAM overlayed predictions of cyclonic events\\
 (a) TC KYAAR (VSCS,center), and MAHA (D,bottom right) at 30 OCT 2019, 0600 hours, AS.\\
 (b) TC MOCHA (ESCS) at 13 MAY 2023, 1100 hours, BOB.}
	\label{fig:subfigures}
\end{figure}

With this motivation, we utilized Eigen-CAM to verify the features which the model was focusing on while making the predictions. The predicted bounding boxes overlaid on top of the activation map for two test instances, one for a multicyclonic instance (KYAAR,MAHA) from 2019  and for cyclone MOCHA in 2023 is depicted in Figure~\ref{fig:subfigures}, highlighting the important features used while making the predictions.

\subsection{RESULT AND ANALYSIS}
In our study, we prototype and run various state-of-the-art classification models using Python and we use Pytorch~\cite{b12}, and test out new architectures and also to customize the architectures based on our requirements,  we used an Nvidia 4090 GPU, with 450 W power consumption, due to its computational abilities, essential for the training and testing of various neural network models. This GPU is equipped with 24 GB of usable graphic memory, which is crucial for managing large-scale data and complex calculations efficiently.
\subsubsection{\textbf{TC Geolocation Detection}}

\begin{figure}[!ht]
    \centering
    \includegraphics[width=1\linewidth]{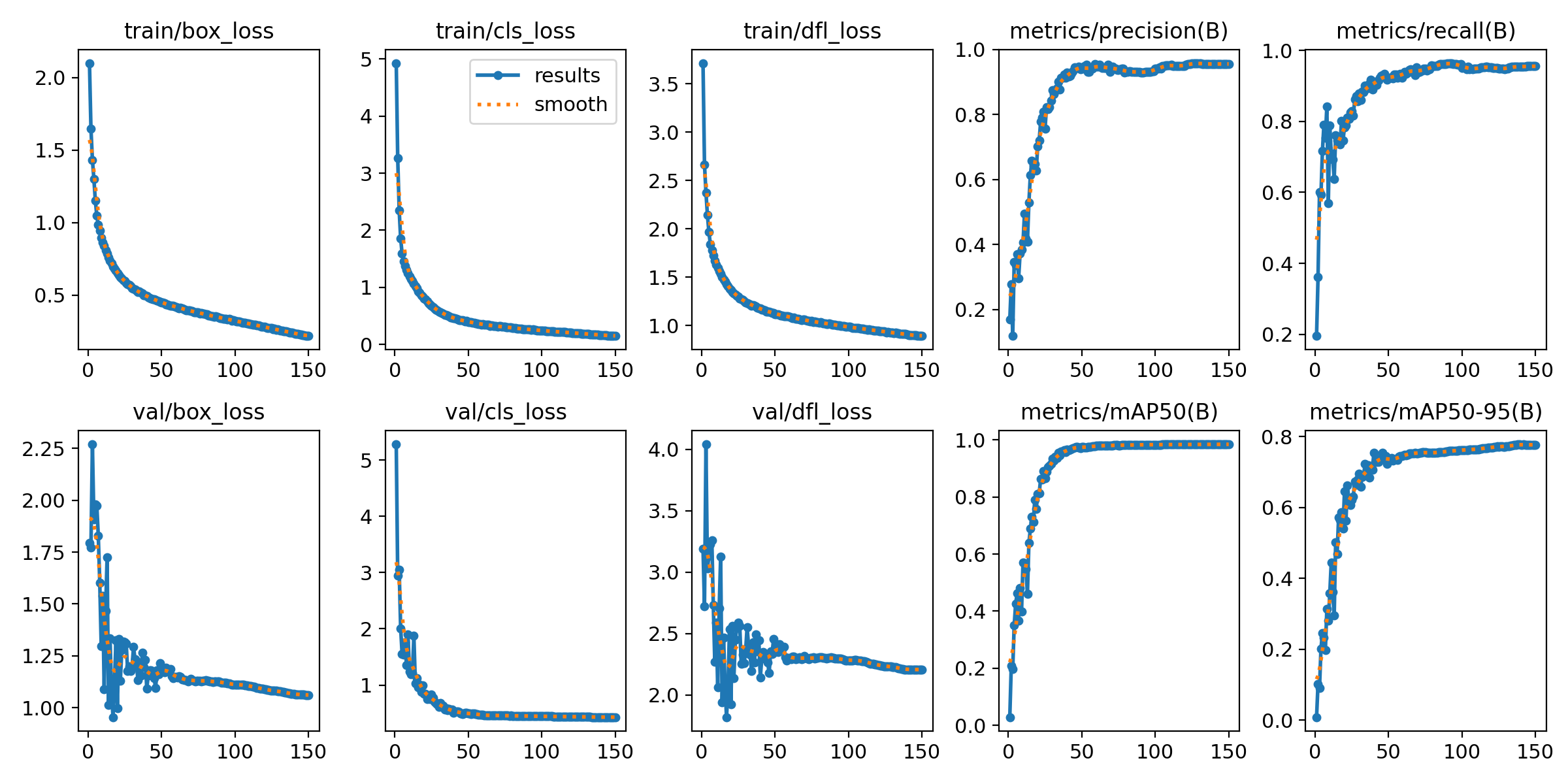}
    \caption{Training and Validation Plots for TC Detection Model}
    \label{fig:yolo_loss_plots}
\end{figure}

The test set comprises of cyclonic data from year 2023 (1828 images), Multicyclonic event data from 2019 (TC MAHA and KYAAR) (130 images) and SuCS event from year 2020 (43 images), to cover all possible scenarios. This data was not a part of the train and val sets. The train and val dataset size was 12709 and 2244 images respectively, spanning cyclones from 2013-2022.All the variants were trained till 150 epochs for consistency. The loss plots of training and validation sets for the Final YOLOv8 model is shown in Figure~\ref{fig:yolo_loss_plots}. The confidence threshold was kept as 0.25 during inference. 
We have also used the NC data (as discussed in the Dataset section) for evaluating our models ability to differentiate between Cyclone vs No Cyclone, and monitor the number of false positive(FP).
Apart from NC FP, in cyclonic images there were Background FP. These are filtered out based on two criteria. 1.) Both Pixel error in X and Y from ground truth is greater than 100 pixels 2.) The model predicted multiple boxes around the same cyclone, which was due to prediction of different classes. A custom threshold of 10 pixel distance error was kept to filter out duplicate predictions in such cases, and higher confidence prediction was retained. The other is termed as a Background FP case.

\paragraph{\textbf{Study I: Variation of Box sizes with TC Intensity}}\label{AA}
The first study was to compare the results according to variation of box sizes during training. The test results for the different configuration along with the customised configuration are tabulated in Table ~\ref{tab:config_comparision}.

\begin{table}[]
\resizebox{\columnwidth}{!}{%
\begin{tabular}{|l|l|l|l|}
\hline
\textbf{Box Size Configuration}     & \textbf{300x300} & \textbf{Custom}    & \textbf{500x500} \\ \hline
\textbf{Avg Km Error}           & 54.45            & 64.18              & 65.28           \\ \hline
\textbf{Avg Lat Long Error}     & (0.323, 0.324)   & (0.358,0.395)      & (0.39, 0.373)   \\ \hline
\textbf{SC No Detections (FN)}      & 317/1871         & 176/1871           & 130/1871         \\ \hline
\textbf{False Positives NC Type I}  & 366/1499         & 436/1499           & 637/1499         \\ \hline
\textbf{False Positives NC Type II} & 168/1500         & 181/1500           & 247/1500         \\ \hline
\textbf{Recall (TP/TP + FN)}        & 0.83            & 0.905              & 0.93            \\ \hline
\textbf{Precision (I,II) NC}        & 0.81,0.90        & 0.79,0.90          & 0.73,0.87        \\ \hline
\textbf{F1-score (I,II) NC}         & 0.82,0.86        & \textbf{0.85,0.90} & 0.82,0.90        \\ \hline
\textbf{Test mAP50,50-95}           & 0.3,0.112        & 0.342,0.253        & 0.291,0.132      \\ \hline
\end{tabular}%
}
\caption{Metrics for varied Box-Size Configurations}
\label{tab:config_comparision}
\end{table}

The lat-lon error observed for the customised configuration, was 0.358,0.395 respectively. Allthough recall (0.93) was highest for 500x500 box size compared to custom (0.905), but the precision was much lower. This was consistent with the fact that most of the no detections were also in the case of D,DD intensity class, where cloud structures are not compact. We observed that for lower box sizes, although detections were lower, but more accurate for higher intensity classes. So levaraging this information we designed the custom configuration, with reduced box sizes to 400 x 400 for (D,DD,CS,SCS,VSCS) , we are able to avoid these False Positives which were primarily from D,DD classes. For ESCS,SuCS we achieved further improvement in positional error when box sizes are reduced to 300 x 300. Thus leading to the optimum box size configuration. 
In case of 500 x 500 configuration, background FP cases observed were 44 compared to 30 in 300x300 configuration and 24 in custom. Overall F1 score of the custom variant was highest, if additionally tested NC data is used for calculation of precision. If the precision is calculated over only cyclonic instances, it amounts to 98.6\% for the custom configuration. While reporting the km error we have used the Haversine formula, to account for the earth's curvature. 

\paragraph{\textbf{Study II: Performance Across TC Intensity Classes}}\label{AA}

\begin{figure}[!ht]
    \centering
    \includegraphics[width=1\linewidth]{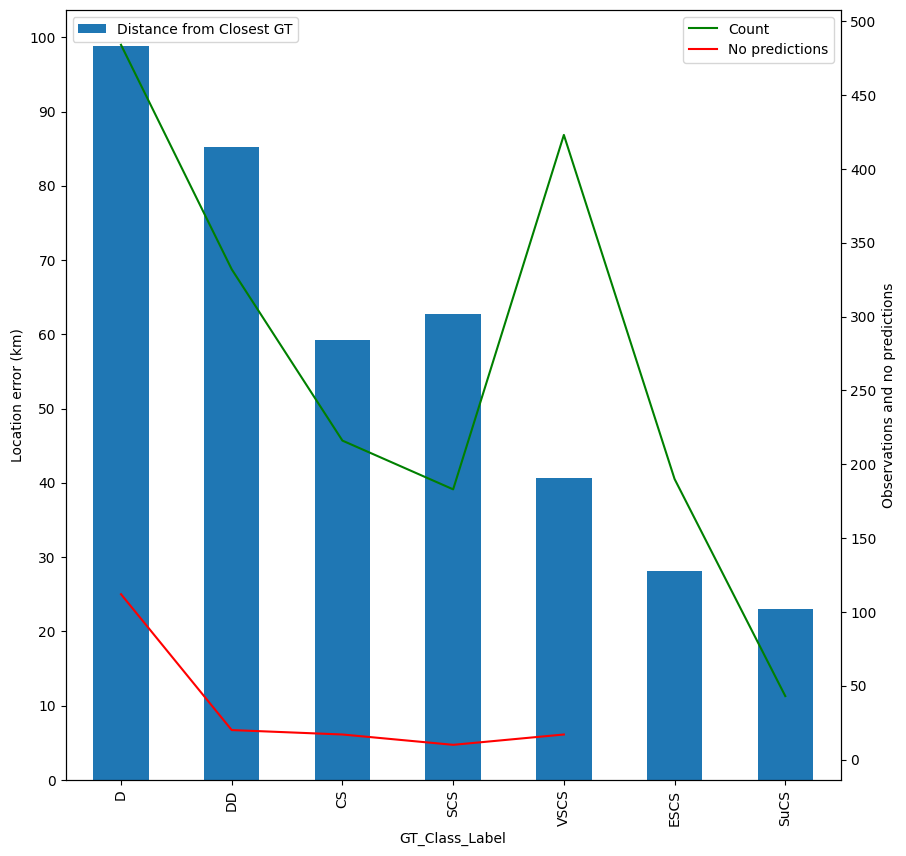}
    \caption{Location errors (y-axis) variation with Intensity Classes (x-axis) , along with no predictions (red) and total test counts (green).}
    \label{fig:class_wise_bar_chart}
\end{figure}

\begin{figure}[!ht]
    \centering
    \includegraphics[width=1\linewidth]{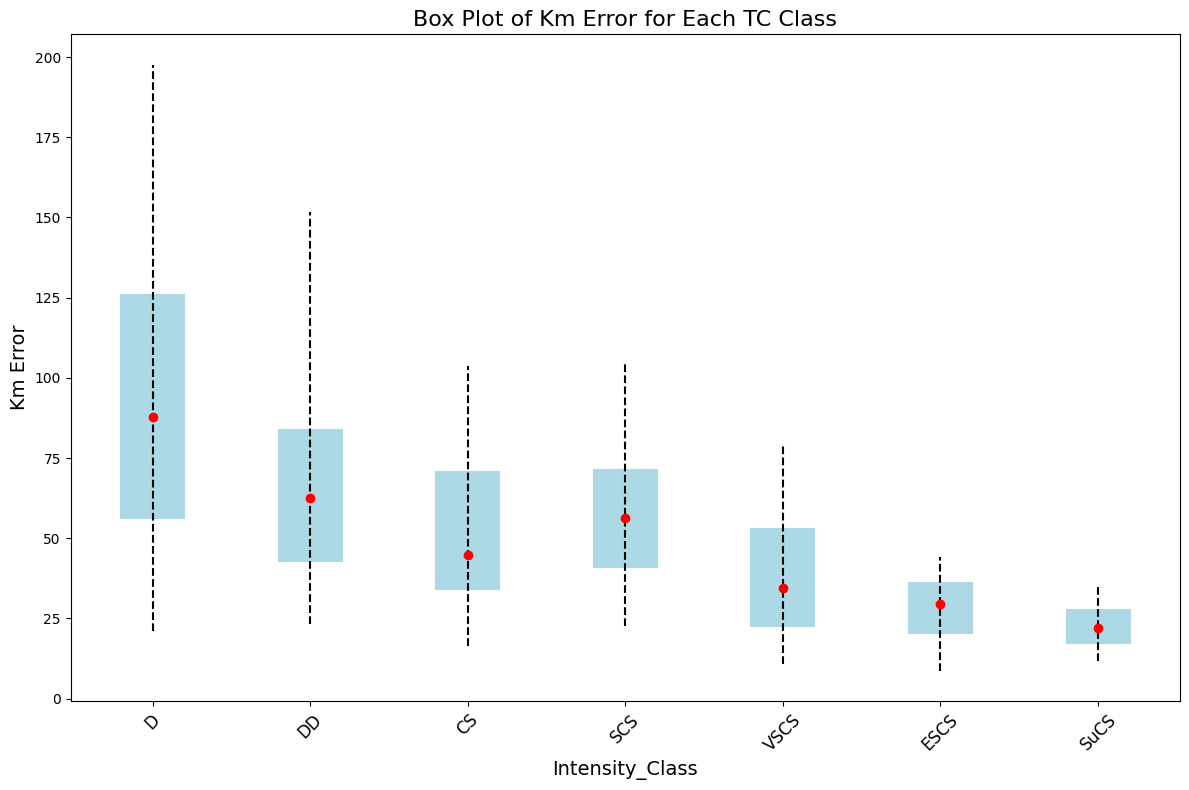}
    \caption{Classwise Box-Plots: Inner box cover the 30th–70th and outer line covers the 10th–90th-percentile error values, of samples, red dot represents median error.}
    \label{fig:box_plot}
\end{figure}

The classwise performance breakdown for the test data (excluding the multicyclonic event) for the custom configuration is tabulated in Table ~\ref{tab:Classwise_Error}. Notably, the error in lat-lon estimation for the lower intensity classes is the highest, as the system is not fully developed and it is difficult to locate the TCs , and decreases significantly for the highest intensity class SuCS. The findings are consistent with that of Malakar P. et al. ~\cite{b18}, which discusses the performance from several reanalysis datasets, over the NIO for TCs from 2008-2015, showcasing similar trends for the errors with increasing intensity. Also, almost 75\% of the total False Negatives are accounted by D, DD classes, which demonstrates even detection is difficult in these circumstances. The same is captured in the Figure~\ref{fig:class_wise_bar_chart}. The box plots showing for each TC class is also represented in the Figure~\ref{fig:box_plot}.

\begin{table}[]
\resizebox{\columnwidth}{!}{%
\begin{tabular}{|l|l|l|l|l|l|}
\hline
TC Class & \textbf{Lat Error} & \textbf{Lon Error} & \textbf{KM error} & \textbf{FN} & \textbf{Count} \\ \hline
\textbf{D}       & 0.557 & 0.595 & 98.76 & 112(63.4\%) & 484  \\ \hline
\textbf{DD}      & 0.454 & 0.530 & 85.209  & 20 (11.4\%) & 332  \\ \hline
\textbf{CS}      & 0.338 & 0.370 & 59.292 & 17(9.7\%)   & 216  \\ \hline
\textbf{SCS}     & 0.308 & 0.427 & 62.732 & 10(5.7\%)   & 183  \\ \hline
\textbf{VSCS}    & 0.242 & 0.239 & 40.603 & 17(9.8\%)   & 423  \\ \hline
\textbf{ESCS}    & 0.166  & 0.169 & 28.089  & 0 (0\%)     & 190  \\ \hline
\textbf{SUCS}    & 0.151 & 0.112 & 22.945  & 0 (0\%)     & 43   \\ \hline
\textbf{Overall} & 0.358    & 0.395    & 64.18     & 176         & 1871 \\ \hline
\end{tabular}%
}
\caption{Classwise Error Performance on Test Set}
\label{tab:Classwise_Error}
\end{table}

\paragraph{\textbf{Study III: Performance on Multi-Cyclone Case in same Basin}}\label{AA}

To demonstrate that the model is capable of detecting multiple cyclones, the event 7 Cyclone KYAAR- event 8 Cyclone MAHA from year 2019 (total 130 images) was also evaluated as a test case. Starting 30OCT2019 0000 HRS, there were two cyclones within AS region. KYAAR was already in its dissipating stages, though still at VSCS category, when another storm MAHA started forming in the same basin AS. By 2NOV2019 1800hrs, MAHA had intensified to SCS, whereas KYAAR dissipated completely. We are able to demonstrate the ability to detect multiple cyclones in the same basin, through the current model, even though there was only 1 such other case in the training data. The number of 2 detections, 1 detection and no detections observed were 34,61,35 respectively. The no detections were primarily in the formation stages of MAHA and dissipation stages of KYAAR. One of the predictions overlayed by Eigen-CAM can be seen in Figure~\ref{fig:subfigMC}.

\paragraph{\textbf{Study IV: Performance on NC data}}\label{AA}

The next study we carried out was the breakdown of the false positives. Overall the classwise breakdown of NC data both Type 1 and Type II is observed. We note that the number of false positives in case of Type I (48hr in past or future)  is higher 436, as compared to 181 observed in Type II (48hr-162hr in past or future) . This confirms a temporal relationship of false positives with the nearest TC. This is also consistent with the findings of Gardoll, S. and Boucher, O.,et.al (2022)\cite{b19} which although use ERA5 renalysis dataset cyclone centered images, but report that the false positives were temporally close to a TC by an average of 131h, approximately 5.5d (in the past or in the future).
Out of the False Positives cases 95\% were classified as D,DD class. Only 5\% of total False Positives was classified as CS and above, which can be called as a serious false positive. The results are tabulated in Table~\ref{tab:NC_Results}.

\begin{table}[]
\begin{tabular}{|l|l|l|l|l|}
\hline
& \textbf{Test Data} & \textbf{FP(D,DD)} & \textbf{FP(CS)} & \textbf{Total FP} \\ \hline
\textbf{NC Type I}  & 1499                      & 417(95.6\%)                     & 19(4.3\%)                     & 436(100\%)        \\ \hline
\textbf{NC Type II} & 1500                      & 172(95\%)                       & 9(5\%)                        & 181(100\%)        \\ \hline
\end{tabular}
\caption{Model Performance on NC Data}
\label{tab:NC_Results}
\end{table}
\paragraph{\textbf{Study V: Performance across YOLO series}}\label{AA}

Further we compared the performance for this configuration with YOLOv7. Although outperforming in recall, the overall F1 score of the YOLOv8 model was still better. The results are tabulated in Table~\ref{tab:model_comparision}.
\begin{table}[]
\resizebox{\columnwidth}{!}{%
\begin{tabular}{|l|l|l|}
\hline
\textbf{Metric}               & \textbf{YOLOv7} & \textbf{YOLOv8} \\ \hline
\textbf{Km Error}             & 100.486         & 64.18          \\ \hline
\textbf{Lat-Lon Error}        & (0.626, 1.086)  & (0.363, 0.395)  \\ \hline
\textbf{False Negatives (FN)} & 45/1871         & 176/1871        \\ \hline
\textbf{FP NC Type I}         & 860/1499        & 436/1499        \\ \hline
\textbf{FP NC Type II}        & 374/1500           & 181/1500        \\ \hline
\textbf{Recall (TP/TP + FN)}  & 0.976           & 0.907           \\ \hline
\textbf{Precision (I,II) NC}  & 0.679,0.83     & 0.79,0.90      \\ \hline
\textbf{F1-score (I,II) NC}   & 0.80,0.897     & 0.85,0.90     \\ \hline
\textbf{Test mAP50,50-95}     & 0.251,0.152     & 0.342,0.253     \\ \hline
\end{tabular}%
}
\caption{Test performance across YOLOv7 and YOLOv8}
\label{tab:model_comparision}
\end{table}

\subsubsection{\textbf{TC Intensity Estimation}}
As discussed in the model development section, we have used a standalone regression model, and the CNN-LSTM  time series models. 
The total centric images extracted were 18,267 images from year 2013-2023. Data for the years 2013-2020 and year 2022 was used in training configuration 1, (13851 samples) and year 2021 was chosen for validation (1237 samples). The year 2023 was used as a test year. The total testing dataset size was 1836 samples, out of which 1421 datapoints before landfall were considered for testing.
To eliminate any bias in test results, we carried out a separate split of data, namely configuration 2, with 2019 event in testing (2272), 2021 event in val (1257), and rest years from 2013-2018,2020,2022-2023 events in training (13668).

For SuCS, limited data was available, as there are only two events one in 2019 (KYAAR) and the other in 2020 (AMPHAN). By choosing two splits, we are able to utilise one in training and the other in testing interchangeably.

\paragraph{\textbf{M1: Single Image Regression Model} }\label{AA}

We experimented with 3 sizes of cyclone centric images, trained for 400 epochs. Loss function chosen was MSE Loss; weight decay was set to 0.0001,  and learning rate was set to 0.001. We have used the Adam optimizer. We compared the performance across various image sizes, and found 300 x 300 was the most ideally suited, with regards to error and computation time, and is also in accordance with the average TC size of 300-600km in NIO. The Train-val loss plot for the M1 model is captured in Figure~\ref{fig:subfigA}.

\paragraph{\textbf{M2: Sequential CNN-LSTM Model}}\label{AA}

For the sequential model, we have experimented with different intervals and durations, all trained for 400 epoch duration. Total training sequences formed were 5515 and 5530, for training 1 and 2 configurations respectively. Validation sequences formed were 487, for interval 3hr, duration 6hr and sequence length as 3. Loss used was MSE Loss, Adam optimiser is used and learning rate  was set to 0.0001. We also did moving average smoothing, with window size 6 and it improved the overall RMSE for the M2 model.
The loss plots for the CNN-LSTM model is shown in the Figure ~\ref{fig:subfigB}

\begin{figure}
	\centering
	\begin{subfigure}{0.45\linewidth}
		\includegraphics[width=\linewidth]{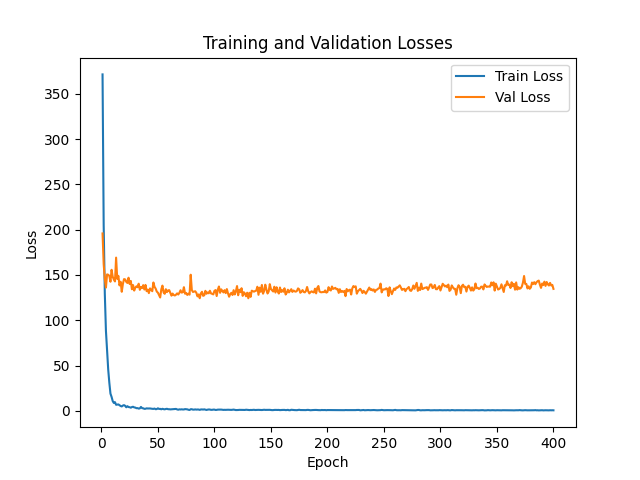}
		\caption{}
		\label{fig:subfigA}
	\end{subfigure} \hspace{0.5mm}
	\begin{subfigure}{0.45\linewidth}
		\includegraphics[width=\linewidth]{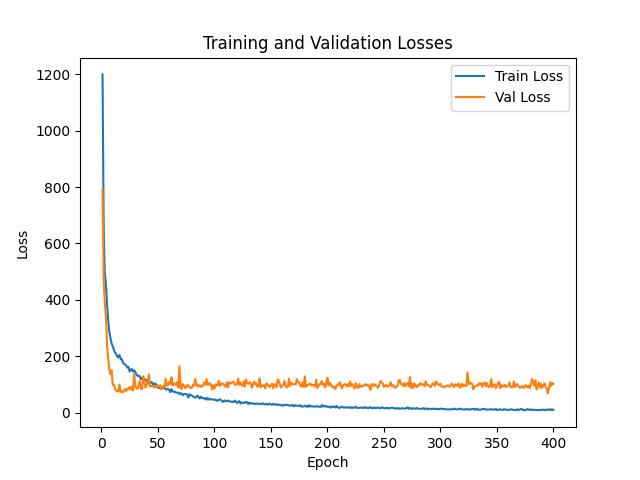}
		\caption{}
		\label{fig:subfigB}
	\end{subfigure}

	\caption{Training and validation loss plots for a) M1 and b) M2 model. }
	\label{fig:subfigures_2}
\end{figure}

The results showing comparison of both the models are tabulated in Table ~\ref{tab:TCI_comparision}.
Comparing all the approaches, the sequential CNN-LSTM M2 model, performed better than the M1 model, for CS,SCS,VSCS classes. Both the models although underestimate the higher intensity classes, ESCS,SuCS. Lack of sufficient number of training datapoints for these scenarios is one of the major cause for this problem. Overall performance was better in the 2023 test case over the 2019 test case, and trends were comparable.

\begin{table}[]
\label{tab:my-table}
\resizebox{\columnwidth}{!}{%
\begin{tabular}{|l|l|l|l|l|l|l|}
\hline
\textbf{Class} & \textbf{Test 2019} & \textbf{M1} & \textbf{M2} & \textbf{Test 2023} & \textbf{M1} & \textbf{M2} \\ \hline
\textbf{D}    & 408 & 9.45  & 9.64  & 303 & 5.7   & 6.42  \\ \hline
\textbf{DD}   & 398 & 8.5   & 5.45  & 228 & 5.66  & 7.67  \\ \hline
\textbf{CS}   & 454 & 14.74 & 8.77  & 145 & 8.5   & 8.15  \\ \hline
\textbf{SCS}  & 297 & 15.96 & 15.64 & 149 & 13.84 & 11.58 \\ \hline
\textbf{VSCS} & 450 & 22.6  & 25.94 & 409 & 32.06 & 19.5  \\ \hline
\textbf{ESCS} & 183 & 33.09 & 41.06 & 187 & 31.71 & 36.4  \\ \hline
\textbf{SuCS} & 82  & 39.99 & 39.93 &  -   &  -    &  -     \\ \hline
\end{tabular}%
}
\caption{Classwise Performance comparision (RMSE) of M1 and M2 models for two different training and test sets.}
\label{tab:TCI_comparision}
\end{table}

\subsubsection{\textbf{Case Studies - Tropical Cyclones over NIO in 2023}}

\begin{figure}
	\centering
	\begin{subfigure}{0.8\linewidth}
		\includegraphics[width=\linewidth]{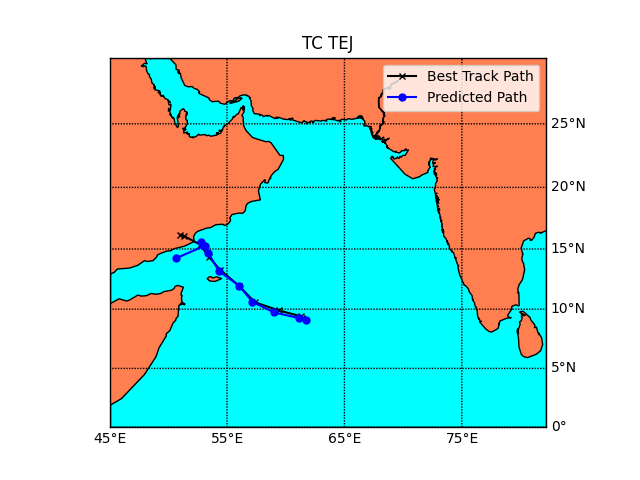}
		\caption{}
		\label{fig:subfigA}
	\end{subfigure} \hspace{0.1mm}
	\begin{subfigure}{0.80\linewidth}
		\includegraphics[width=\linewidth]{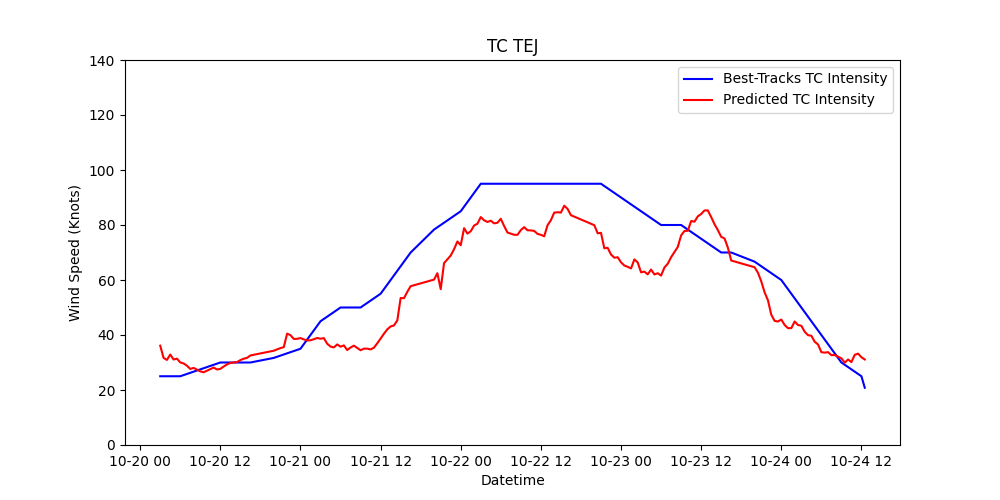}
		\caption{}
		\label{fig:subfigB}
	\end{subfigure}

	\caption{Track (a) and Intensity (b) predictions for TEJ, compared to best tracks.}
	\label{fig:Tej}
\end{figure}

We present next case studies for 3 TCs of different intensities, in year 2023 in the NIO, as tabulated in Table~\ref{tab:TC_CASES}. . 
\begin{table}[]
\caption{Details of cyclones formed over NIO in 2023}
\resizebox{\columnwidth}{!}{%
\begin{tabular}{|l|l|l|l|l|}
\hline
\textbf{No.} & \textbf{Cyclone Name} & \textbf{Life Duration} & \textbf{Max Intensity} & \textbf{Basin} \\ \hline
1 & Tej      & 20-24 Oct 2023 & ESCS (95 Kts) & AS  \\ \hline
2 & Hamoon   & 21-25 Oct 2023 & VSCS (65Kts)  & BOB \\ \hline
3 & Michaung & 01-06 Dec 2023 & SCS (55Kts)   & BOB \\ \hline
\end{tabular}%
}
\label{tab:TC_CASES}
\end{table}
\paragraph{\textbf{Cyclone TEJ}}\label{AA}
TEJ was the sixth cyclonic event of 2023, which IMD started monitoring from 16th Oct 2023, in central south AS region. It coexisted with Cyclone Hamoon in BOB. The cyclone intensified to ESCS (Extremely Severe Cyclonic Storm,95 Kts) category by Oct 22, and made landfall over Yemen. The Track and intensity predictions using the modules developed is shown in the Figure ~\ref{fig:Tej}. The mean track error from our model was 36.5 km and the RMSE and MAE in intensity was 12.47, 10.54 knots respectively. 

\begin{figure}
	\centering
	\begin{subfigure}{0.80\linewidth}
		\includegraphics[width=\linewidth]{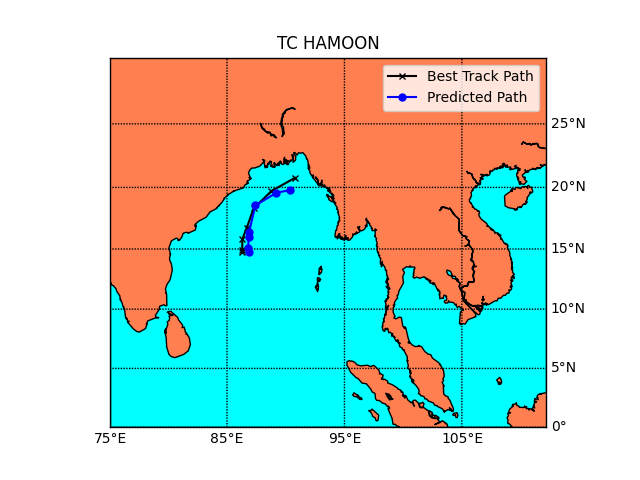}
		\caption{}
		\label{fig:subfigA}
	\end{subfigure} \hspace{0.1mm}
	\begin{subfigure}{0.80\linewidth}
		\includegraphics[width=\linewidth]{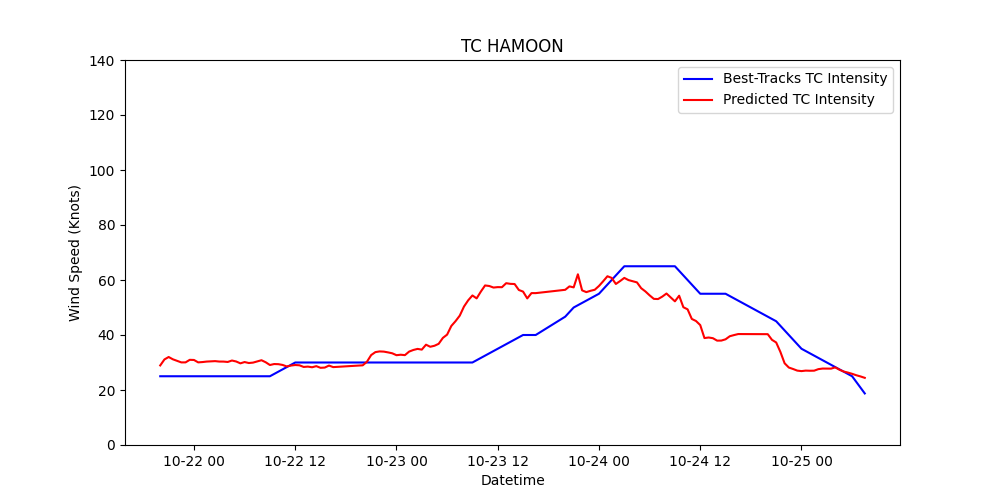}
		\caption{}
		\label{fig:subfigB}
	\end{subfigure}

	\caption{Track (a) and Intensity (b) predictions for HAMOON, compared to best tracks.}
	\label{fig:Hamoon}
\end{figure}

\paragraph{\textbf{Cyclone HAMOON}}\label{AA}
On 21st Oct, there was a low-pressure formation in BOB, which intensified into VSCS (Very Severe Cyclonic Storm, 65Kts) category by 24th Oct, named HAMOON subsequently. It crossed Bangladesh coast, and dissipated over Myanmar.
Track and intensity predictions using the modules developed is shown in the Figure ~\ref{fig:Hamoon}.
The mean track error from our model was 65.15 km and the RMSE and MAE in Intensity was 10.17, 7.95 knots respectively.

\begin{figure}
	\centering
	\begin{subfigure}{0.80\linewidth}
		\includegraphics[width=\linewidth]{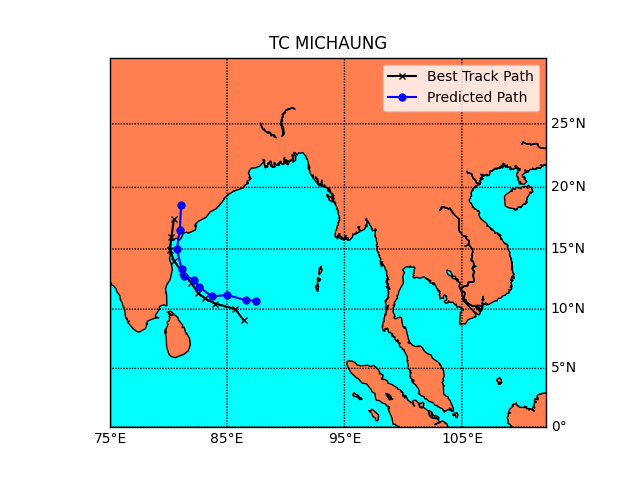}
		\caption{}
		\label{fig:subfigA}
	\end{subfigure} \hspace{0.5mm}
	\begin{subfigure}{0.80\linewidth}
		\includegraphics[width=\linewidth]{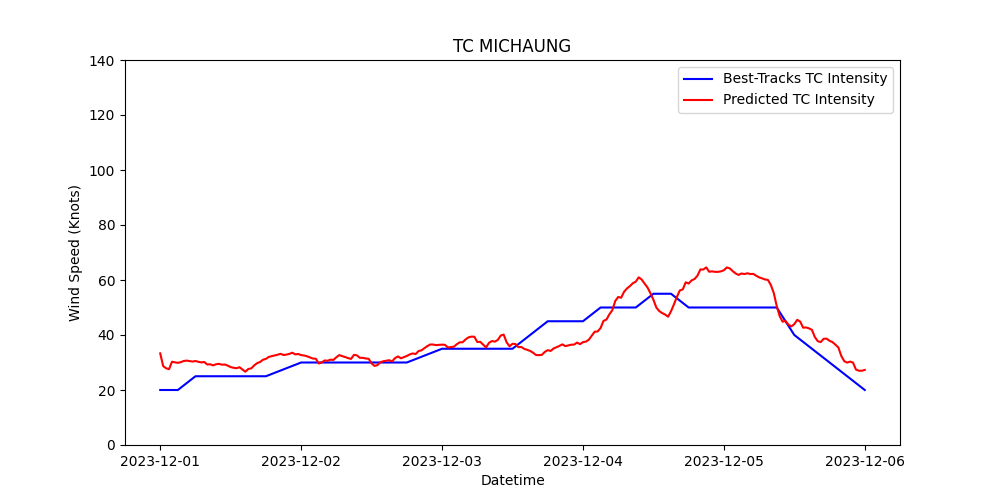}
		\caption{}
		\label{fig:subfigB}
	\end{subfigure}

	\caption{Track (a) and Intensity (b) predictions for MICHAUNG, compared to best tracks.}
	\label{fig:Michaung}
\end{figure}

\paragraph{\textbf{Cyclone MICHAUNG}}\label{AA}
MICHAUNG originated in the Gulf of Thailand, crossing into BOB. IMD starting tracking it from 1st Dec. The TC intensified to a SCS (Severe Cyclonic Storm, 60knots) by 4th Dec, and gradually moved towards the east coast of India, causing heavy rainfall over Tamil Nadu and Andhra Pradesh. It made landfall in Andhra Pradesh on 5th Dec.
The Track and the intensity predictions using the modules developed is shown in the Figure ~\ref{fig:Michaung}.
The mean track error from our model was 85.29 km and the RMSE and MAE in Intensity was 6.62, 5.47 knots respectively.

\section{Results Comparison with Related Works} 
1. As per our best knowledge, this is the first work that discusses effect of variation of box sizes across intensity classes over the NIO, and utilises it for enhancing object detection performance.\\ 
2. We achieve the best overall lat-lon error of (0.323, 0.324) which outperforms the work by T. Long et al.~\cite{a4}, with reported lat-lon error of (0.41,0.37) over TCs forming over northwest Pacific Ocean, although they consider 6 channels. We also outperform the classwise lat-lon errors for the intensities CS and above compared to their model. Wang P . et al.~\cite{a5} are able to demonstrate mean lat-lon error of 0.237, using DL+IP approaches, but does not consider TD class intensity, in overall error calculations.\\
3. There is limited literature with explicit use of NC data, for evaluating the model performance for object detection. D.Galea et al. (2023) ~\cite{b20} uses ERA-Interim reanalysis data from 1979 to 2017 to determine TC presence. They tested using data from two subsequent years, and obtained a recall rate of 92\%. The precision rate obtained was that of 36\%, although it consisted data from all oceans of the world. We demonstrate a precision of 79\% for NIO region only, over 48 hr data in future or past from a TC, which further improves as temporal distance from a TC instance increases.\\
4. Objective Detection of Center of TC in Remotely Sensed IR Images (Neeru Jaiswal et al.(2013)~\cite{a1}), discusses track performance across 4 cyclones over NIO, achieving track errors of 42, 82, 58, 42.5 km. The tracking is done over centered images, which decreases the error inherently,using a guess center approach. We demonstrate a track performance of 36.5, 65.15, 85 km over 3 cyclones discussed as case study, and overall 64 km track error, without the need of determining a guess center which has to be estimated manually using BT values and could lead to FP.\\   
5. Cyclone intensity estimation using similarity of satellite IR images based on histogram matching approach , Neeru Jaiswal et al.(2012)~\cite{b21}), reports MAE for intensity less than 34 kt as 11.84 , 34-63 as 8.81, 64-95 as 14.51, 95 and above as 19.60kt. We are able to outperform MAE for all the classes till VSCS category, being 4.92(D), 5.14(DD), 6.61(CS), 10.44(SCS), 17.79 knots (VSCS) respectively for 2023.\\ 
6. With regards to intensity estimation, Zhang C.J. et al.(2020)~\cite{b4} uses a classifier regression based approach, but does not consider lower intensity data (TD), and achieves an RMSE of 8.25 for TS class. We establish benchmark performance for D, DD, CS classes as 6.47, 7.67, 8.15 knots respectively in comparison.\\

\section{Future Work}
The current work is able to establish the necessary modules to build a robust pipeline to detect the location of cyclone and thereafter estimate intensity across the NIO region.
Future work will include increasing the robustness of the object detection model, including augmentations. Further, time series aspects are currently incorporated only in the intensity estimation module, which could also be extended in cyclone detection. Though the current model is capable of identifying multiple cyclones as demonstrated, for increasing robustness, artificially generated samples with multiple cyclones could also be utilised in future for training. The current errors in cyclone detection for the lower intensity classes could be further reduced by incorporating a refinement model, which would enable further correction over centric images.
With regards to intensity estimation module, the current models underestimate the samples from higher intensity classes, especially ESCS, SuCS. Generative modelling, could also be explored apart from standard augmentation techniques, and data from other oceans could be incorporated to deal with this problem. Further, we observed that, behaviour of a TC changes rapidly when it approaches landfall, making intensity estimation even more difficult. Separate Post landfall intensity models, like discussed in the work by Kishtewal et al. (2013)~\cite{b19}, have to be incorporated to overcome these limitations.

\section*{Acknowledgment}
We extend our gratitude to ISRO SAC (Space Applications Center) Ahmedabad for their support and MOSDAC (Meteorological and Oceanographic Satellite Data Archival Centre) for provision of INSAT-3D IR Image data along with invaluable technical expertise. We are grateful to the ISRO-IISc STC (Space Technology Cell) and ARTPARK IISc for their support in facilitating this research.

\vspace{20pt}
\end{document}